# Structural modulation in potassium birnessite single crystals

Liliia D. Kulish,*[a] Pavan Nukala,[b] Rick Scholtens,[a] A. G. Mike Uiterwijk,[a] Ruben Hamming-Green[a] and Graeme R. Blake[a]

[a] Zernike Institute for Advanced Materials, University of Groningen, Nijenborgh 4, 9747 AG, Groningen, the Netherlands. E-mail: l.kulish@rug.nl
[b] Center for Nanoscience and Engineering, Indian Institute of Science, Bengaluru, 560012, India.

We report on the growth of single-crystal potassium birnessite ($K_{0.31}MnO_2 \cdot 0.41H_2O$) and present both the average and local structural characterization of this frustrated magnetic system. Single crystals were obtained employing a flux growth method with a $KNO_3$ / $B_2O_3$ flux at 700 °C. Single-crystal X-ray diffraction revealed an average orthorhombic symmetry, with space group *Cmcm*. A combination of high angle annular dark field scanning transmission electron microscopy (HAADF-STEM) with atomic resolution energy dispersive X-ray spectroscopy (EDS) demonstrated the layered structure of potassium birnessite with manganese-containing planes well separated by layers of potassium atoms. $MnO_6$ octahedra and the K / $H_2O$ planes were clearly imaged via integrated differential phase contrast (iDPC) STEM. Furthermore, iDPC-STEM also revealed the existence of local domains with alternating contrast of the manganese oxide planes, most likely originating from charge ordering of $Mn^{3+}$ and $Mn^{4+}$ along the *c*-axis. These charge-ordered domains are clearly correlated with a reduction in the *c*-lattice parameter compared to the rest of the matrix. The insight gained from this work allows for a better understanding of the correlation between structure and magnetic properties.

## Introduction

Compounds with the birnessite structure have been an object of interest for researchers in different fields since the 1950s.[1] Birnessite compounds consist of 2D planes of $MnO_6$ octahedra separated by ~7 Å with mono- or divalent ($A^+$) cations and water molecules between the layers, with mixed-valent Mn adopting the 3+/4+ oxidation state.[2,3] The general chemical formula is $A_xMnO_2 \cdot yH_2O$ ($A$ = alkali or alkaline earth metal), where the $A$/Mn molar ratio is less than 0.7.

The high mobility of the interlayer cations together with the low toxicity and cost of birnessite materials provide applications in the field of energy storage, such as capacitors showing high cycling capability,[4-6] and as promising cathodes for batteries.[7] Moreover, birnessites have been used as starting materials for further chemical transformations. For example, single crystals of potassium birnessite were used as a precursor for the investigation of protonation, exfoliation and self-assembly of its layered structure.[8] Furthermore, the interlayer space of birnessites allows both the intercalation of different cations by various synthetic strategies[4,9,10] and ion-exchange reactions. For example, the exchange of $K^+$ for $Fe^{3+}$ ions in potassium birnessite was investigated in a study of photocatalytic water oxidation.[11] Birnessite compounds have also been demonstrated as materials for the extraction of radioactive $Sr^{2+}$ from nuclear waste[12] and as molecular sieves for purposes such as water purification.[13]

In contrast, relatively few studies have been reported on the magnetic properties of birnessites.[14] Accumulated experience from other layered materials provides a solid foundation for starting an investigation of the potentially unique magnetic behavior of birnessite compounds, where magnetic and non-magnetic layers alternate. Modifying the interlayer distance and the composition of the interlayer species by different topotactic reactions can affect the ratio of $Mn^{3+}$/$Mn^{4+}$, allowing tuning of the magnetic exchange and anisotropy along the stacking direction. Moreover, as the magnetic sheets are comprised of triangular arrangements of manganese cations, there are conflicting antiferromagnetic interactions that cannot be mutually satisfied, giving rise to magnetic geometrical frustration.[15-17] Geometrical frustration promotes the existence of a manifold of ground states that are close in energy, often leading to spin liquid and spin ice behavior and the emergence of exotic magnetic states, such as helical and cycloidal spirals or even a variety of periodic states with non-trivial topologies composed of skyrmions and antiskyrmions.[18-20] This prospect makes the study of the magnetic nature of birnessites particularly valuable. A new look at these well-known compounds from a different perspective may add a further chapter to their investigation and open a new field of interest.

Recently, we reported one of the first investigations of the magnetic properties of polycrystalline birnessite compounds. Cluster glass behavior was revealed below the freezing temperature of 4 K for sodium birnessite and 6 K for potassium birnessite.[14] Only one earlier study had been performed, on water-containing sodium birnessite, but this investigation was complicated by the presence of impurities of other magnetic phases.[21] No information is available on the magnetic properties of other $A_xMnO_2 \cdot yH_2O$ birnessites.

One of the next steps required in order to gain more insight into the magnetic behavior is the investigation of birnessite single crystals, which will allow anisotropic magnetic properties to be measured.

The current study is dedicated to starting along this path via the growth of potassium birnessite single crystals. Potassium birnessite ($K_xMnO_2 \cdot yH_2O$, $x<1$ hereafter referred to as K-bir) is characterized by layers of edge-shared $MnO_6$ octahedra, which are rotated with respect to each other by 180 degrees such that there are two layers per unit cell along the *c*-axis. The planes are separated by gaps containing $K^+$ cations and $H_2O$ molecules. The interlayer species are located in the prismatic cavities, where it was reported that the $K^+$ ions are shifted within the *ab* plane from the center of the prism toward the three lateral faces; there are thus two different chemical environments.[7,22] A number of different polytypes have been identified, depending on the precise alkali and water content. For example, charge ordering of $Mn^{3+}$ and $Mn^{4+}$ ions can occur, giving rise to an in-plane super-periodicity.[23]

K-bir can be obtained by various chemical methods.[24] Bulk samples of K-bir have been prepared by thermal decomposition of $KMnO_4$ in air,[10,23] by a facile solid state reaction of $K_2CO_3$ and $MnO_2$ after which the proportion of $K^+$ ions can be increased by ion exchange processes,[7] and by the oxidation of $Mn(OH)_2$ in the presence of an excess of the alkali metal.[12] Layered nanoflakes of K-bir have been obtained via hydrothermal heating of potassium oxalate with potassium permanganate aqueous solution.[25]

The studies referred to above all involved polycrystalline samples of K-bir. In general, both natural and synthetic birnessite compounds tend to form as particles of less than 100 nm in size and often much smaller. The growth of these particles can be achieved by further chemical reactions involving changes in cation content, hydration, and oxidation state.[24]

Here, we demonstrate the direct single-crystal growth of K-bir under different synthesis conditions. The largest single crystals obtained are $250 \times 250 \times 20$ μm$^3$. The chemical compositions, uniformity of the compounds, and oxidation state of the manganese atoms are determined by different analytical methods. For the first time, structural characterization of K-bir is carried out using a combination of single-crystal XRD and high resolution scanning TEM. We find that the crystals adopt an average orthorhombic symmetry with space group *Cmcm*. On the local scale, more complex features of the K-bir layered structure are observed, such as stacking disorder, modulation of the structure and the presence of nanoscopic domains in which there is evidence of $Mn^{3+/4+}$ charge order along the c-axis. All of these observations are relevant in obtaining a better understanding of the relations between the structure and magnetic properties.

## Experimental details

### Preparation of K-birnessite

A flux growth method was used for the preparation of K-bir single crystals. Different synthesis routes were attempted. The first method was adapted from an earlier report by Yang *et al.*,[26] in which a 5:1 molar ratio of $KNO_3$ (3.2 g, 0.0317 mol) and $Mn_2O_3$ (1 g, 0.0063 mol) was ground with 1 wt% of $B_2O_3$. The large excess of $KNO_3$ was used as a flux. The addition of $B_2O_3$ leads to a less viscous flux, promoting higher atom mobility and therefore a higher rate of crystallization. The mixture was placed either into a 25 mL alumina crucible and covered with an alumina cap (sample 1) or into a 5 mL platinum crucible with a platinum cap (sample 2). In both cases the crucible was heated to 700 °C over a period of 6 hours and held for 60 hours. Cooling to room temperature was then carried out over a period of 12 hours. The obtained solid (dark brown crystals with dark brown powder) was washed with distilled water to remove the flux and any by-products until the supernatant became colorless and the pH was about 7. Finally, the solid was dried at 60 °C for 10 hours.

Another single crystal growth method used a different flux. A 1.1 : 1 molar ratio of $KNO_3$ : $Mn_2O_3$ was used (0.7044 g of $KNO_3$ and 1 g of $Mn_2O_3$), which was added to a flux of PbO (0.7 g) and $B_2O_3$ (0.7 g) in a 5 mL platinum crucible with cap (sample 3). The same temperature treatment was used as in the previous attempts. The obtained dark powder was separated from the solidified flux by leaching the crucible in dilute nitric acid for several hours. The sample was then washed with distilled water until the pH was 7 and dried at 60 °C for 10 hours.

### Methods of analysis

The average oxidation state of manganese was determined using an oxalic acid – permanganate back titration method.[27] This was supported by X-ray photoelectron spectroscopy (XPS), for which the samples were prepared under ambient conditions and pressed onto carbon tape. An Al Kα X-ray source (VG Microtech Mark 2 twin anode with a VG Microtech Clam 2 hemispherical electron energy) and pass energies of 10 eV (Mn 3s) and 20 eV (Mn 3p) were used, with a resolution of 0.2 eV. The dehydration processes occurring on heating were investigated by means of simultaneous thermogravimetric analysis (TG) and differential scanning calorimetry (DSC) on a TG 2960 SDT instrument using an argon flow of

100 mL/min; the heating rate was 10 °C/min over the temperature range 30 °C to 900 °C. To establish the morphology of the single crystals, as well as the elemental distribution, scanning electron microscopy together with energy dispersive X-ray spectroscopy (SEM with EDS detector, Fei NovaNanoSEM 650) was performed.

The phase purity and the crystal structure of the products were determined by single-crystal and powder X-ray diffraction (XRD). Single-crystal XRD was carried out on a Bruker D8 Venture diffractometer equipped with a Triumph monochromator and a Photon100 area detector, operating with Mo Kα radiation. The crystals were mounted using a 0.3 mm nylon loop and cryo oil. The samples were cooled to 100 K with a nitrogen flow from an Oxford Cryosystems Cryostream Plus. Data processing, structure solution, and refinement were done using the Bruker Apex III software. Powder XRD was performed using a Bruker D8 Advance diffractometer operating with Cu Kα radiation in the 2θ range 10 – 70 °. The powder XRD data were fitted by Rietveld refinement using the GSAS software.[28]

Local structural analysis was performed via various complementary aberration-corrected high-resolution transmission electron microscopy (TEM) techniques. TEM lamellae were prepared by focused ion beam (FIB) -based processing with the crystallographic $a*$ axis (perpendicular to $a$ in a hexagonal setting) as the zone-axis (Fei Helios G4 CX). Energy dispersive X-ray spectroscopy (EDS), high angle annular dark field scanning transmission electron microscopy (HAADF-STEM) and differential phase contrast (DPC)[29-33] imaging were performed on these samples using a Thermofisher Themis-G microscope operating at 300 kV accelerating voltage.

## Results and discussion

### Single crystal growth

Fig. 1 shows SEM images of the products obtained by the different flux growth techniques. The largest single crystals were obtained using the $KNO_3$ – $B_2O_3$ flux and a 25 mL alumina crucible (sample 1) and were larger than in previous reports.[26,34] These crystals have a hexagonal plate-like morphology with a size of up to ~250 μm across the hexagonal surface (Fig. 1a). When viewed in the perpendicular direction, a lamellar structure is observed (Fig. 1b). The crystals have a maximum thickness in this direction of ~20 μm.

Using a 5 mL Pt crucible (sample 2) yielded single crystals of similar morphology but with a much smaller size of 2 - 9 μm across the hexagonal surface. Some crystals have a prismatic shape, suggesting a different crystal growth process (Fig. 1c). Powder XRD showed only the birnessite phase (details of the composition and structure are given in the Supplementary Information (SI), section 1). According to EDS, both samples contain around 1 at.% of aluminum or platinum, which are evenly distributed over the whole volume of the samples. Apparently, the composition and size of the crucible play some role in the crystallization.

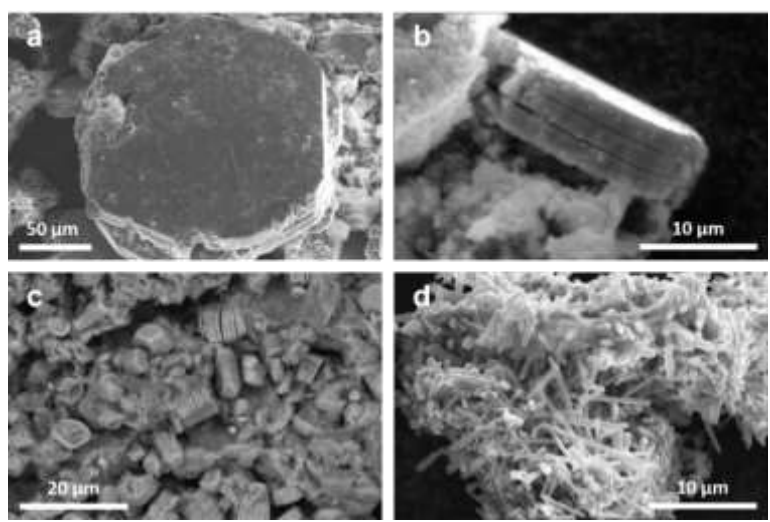

Fig. 1 SEM images of a) K-bir single crystal grown using $KNO_3$ – $B_2O_3$ flux and alumina crucible (sample 1) viewed perpendicular to *a-b* plane and b) perpendicular to *c*-axis; c) sample synthesized using $KNO_3$ – $B_2O_3$ flux and Pt crucible (sample 2); d) sample synthesized with PbO – $B_2O_3$ flux and Pt crucible (sample 3).

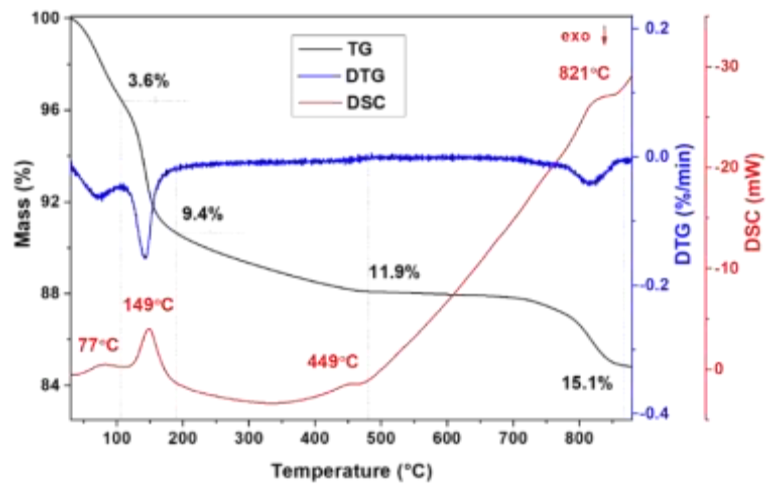

Fig. 2 Thermal analysis of K-bir. The TG and DSC data are represented by the black and red lines respectively.

Synthesis using the PbO – $B_2O_3$ flux (sample 3) yielded a dark powder. SEM images reveal particles with a needle-like shape and a length of 4 - 10 µm (Fig. 1d). However, powder XRD showed that the sample is almost pure cryptomelane ($KMn_8O_{16}$) with the tetragonal space group *I4/m*; there is no trace of any birnessite structure (details are given in the SI, section 2). Cryptomelane contains manganese with a mixed oxidation state of $Mn^{2+}/Mn^{4+}$ and was previously synthesized by the thermal decomposition of $KMnO_4$ at 600 °C.[35] EDS showed the presence of ~1 at.% of lead in the compound. Washing twice with nitric acid and then with water did not remove all the Pb. It was previously shown that if lead is present in higher concentrations when cryptomelane is formed, it can partially occupy the K positions in the structure.[36] Moreover, a phase transformation of K-bir to the cryptomelane structure is possible.[24]

Due to the main goals of the current research involving the characterization of K-bir single crystals, the results and discussion below focus only on the sample obtained using the $KNO_3$ – $B_2O_3$ flux and an alumina crucible (sample 1).

**Determination of chemical composition**

The chemical composition of the K-bir sample was determined by a combination of thermal analysis and a back-titration method. The thermal analysis data are presented in Fig. 2. Four main endothermic effects and corresponding mass losses can be identified in the differential scanning calorimetry (DSC) and thermogravimetric analysis (TG) curves collected on a ground sample heated from 30 – 900 °C under a flow of argon gas. Similar behavior was previously shown for bulk $K_{0.6}MnO_2·0.48H_2O$ with the birnessite structure.[14] The lowest temperature feature corresponds to surface water evaporation with a mass loss of 3.6 % and according to the time derivative of the TG curve (DTG), the loss of water is completed at 97 °C with the maximum of the DSC curve at 77 °C. The second endothermic effect with a sharp maximum in the DSC curve at 149 °C corresponds to the release of interlayer crystal water with a further mass loss of 5.8 %. Further heating leads to the third endothermic effect with the maximum of the DSC curve at 449 °C, shown by a small mass loss of 2.5 %. This corresponds to decomposition of the birnessite structure to $Mn_2O_3$ with the release of oxygen.[37] The last endothermic effect (with a maximum in the DSC curve at 821 °C and a mass loss of 2.7 %) can be attributed to the formation of high-temperature manganese oxide polymorphs.

In the back-titration method of compositional analysis, sodium oxalate was first reacted with the K-bir sample, and the amount of excess oxalate was then determined by titration with $KMnO_4$ solution. Details of the technique are described in Ref. 14. The average manganese oxidation state corresponds to +3.69, assuming no oxygen or manganese vacancies. Thus, 0.31 mol per formula unit of $K^+$ cations are required for charge balance. The amount of interlayer crystal water can then be determined from the TG data as 0.41 molecules per formula unit. This results in a stoichiometry of $K_{0.31}MnO_2·0.41H_2O$.

**Uniformity of the composition**

As described above, the synthesis yielded a mixture of dark brown crystals and powder. Both have the same crystal structure; single crystal XRD data are consistent with powder XRD data collected on the mixture, as discussed below and in the SI, section 7. We note, however, that small fluctuations in composition are impossible to determine by these analytical methods. To check whether the composition is the same over the whole volume of the sample, SEM-EDS analysis was performed on 10 different regions, including the surfaces of single crystals in different orientations and various areas of a pressed powder sample.

It was found that the K:Mn ratio is the same within experimental uncertainty for all regions probed and corresponds to K : Mn = 0.30±0.01 : 1 (details are given in the SI, section 3). This value correlates well with the composition determined by the back-titration method. Furthermore, all elements (K, Mn, O) are evenly distributed over the whole volume of the sample (Fig. 3). As mentioned above, around 1 at. % of aluminum from the crucible was also detected in all the measured regions. No clusters of Al were visible; it was evenly distributed over the whole volume of the sample.

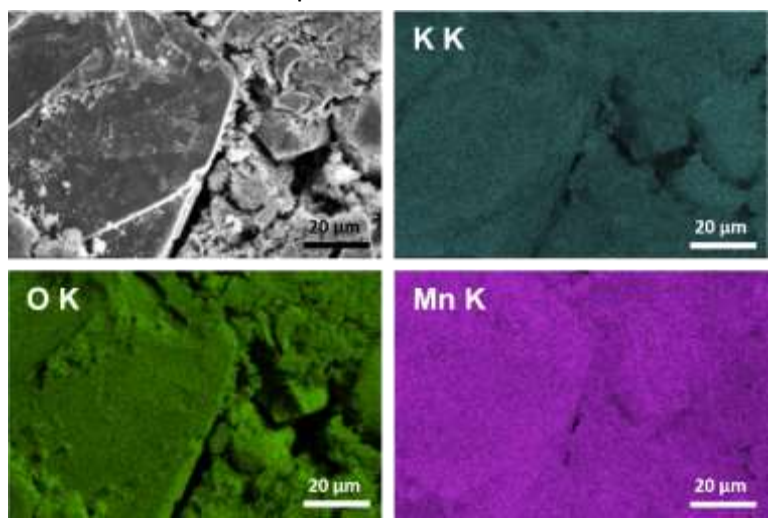

Fig. 3 SEM image (top left) and EDS elemental mapping of composition of K-bir.

## Oxidation state of manganese

In order to determine directly the oxidation state of the Mn atoms in K-bir, XPS was performed focusing on binding energies corresponding to the Mn 3p and 3s states (Fig. 4; an overview of the entire XPS spectrum is presented in the SI, Fig. S8). Due to the exchange interaction between holes and electrons in the 3s and 3d orbitals, the 3s photoelectron peak is split, and is highly sensitive to changes in oxidation state. Typically, it is assumed that there is a linear relationship between the peak splitting and the oxidation state of Mn.[38] However, it has recently been demonstrated that for multivalent layered materials, such as birnessites, this assumption is not reliable.[39] Rather, a comparison with the XPS spectra of samples containing well-known Mn oxidation states is required. Ilton *et al*.[39] used pyrolusite ($MnO_2$) and manganite (MnO(OH)) as standards for $Mn^{4+}$ and $Mn^{3+}$ respectively in determining the oxidation state of Na-birnessites. Despite the higher intensity of the Mn 2p peak, this has previously proven to be unreliable for determining the oxidation state. In addition, the more energetic 3s and 3p photoelectrons are advantageous to use as they originate from deeper in the material.

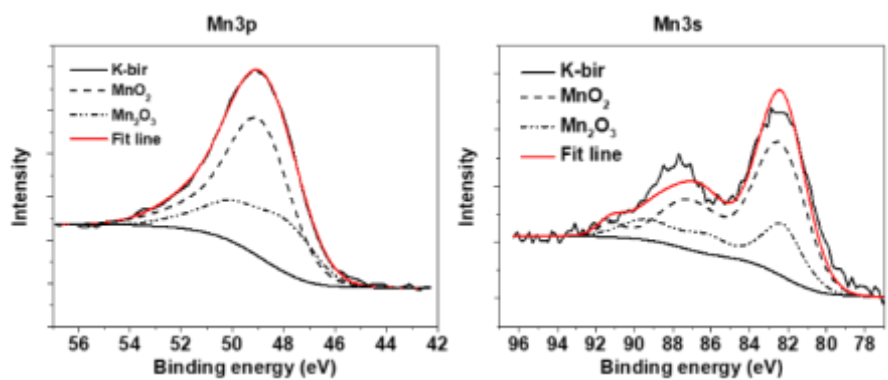

Fig. 4 Fitted XPS spectra (Mn 3p and 3s) of K-bir.

The XPS analysis involves linear addition and scaling of the peaks from each standard in order to best fit the spectrum of a multivalent sample. Additionally, the binding energy of the standard curves is allowed to vary, in order to compensate for differences in the Madelung potential in different compositions/crystal structures. In this work $MnO_2$ and $Mn_2O_3$ powders were used as standards. Shirley backgrounds were used to fit all the spectra, and the reference curves were fitted using a combination of purely Gaussian functions with the RxpsG software.[40] These curves were then fitted to the K-bir peaks, and finally integrated to determine the proportion of Mn(III) and Mn(IV). The fits performed using this method are shown in Fig. 4 and gave an average oxidation state of 3.67+ for the Mn atoms, which corresponds well with the value obtained by the back-titration method.

**Average crystal structure**

For the first time, single-crystal XRD was used to study the structure of K-bir. The diffraction patterns showed evidence of structural modulation, short-range order and/or disorder (see discussion below), but the "average" crystal structure of $K_{0.31}MnO_2 \cdot 0.41H_2O$ could be determined and is shown in Fig. 5; the corresponding crystallographic information file (CCDC 2041937) and structural parameters (Table S7) are available in the SI. Manganese occupies a single crystallographic position, which implies randomly distributed $Mn^{3+}/Mn^{4+}$ cations within the $MnO_6$ layers. The refined Mn-O bond lengths and the Mn-O-Mn bond angles are consistent with an intermediate valence state of $Mn^{3+}/Mn^{4+}$ (see the SI, Table S8). Electron density mapping did not allow the distribution of $K^+$ cations and water molecules to be unambiguously determined in the interlayer space. Therefore, the distribution of $K^+/H_2O$ was taken to be random in the structural model used in the fitting.

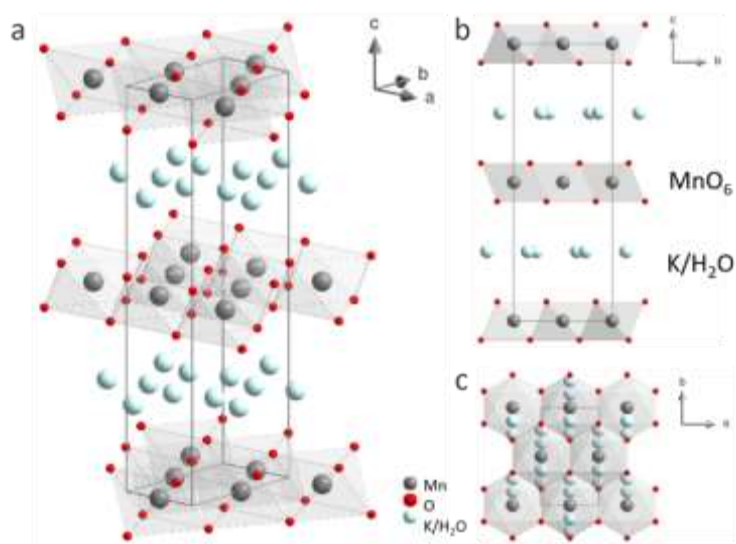

Fig. 5 Schematic representation of the crystal structure of $K_{0.31}MnO_2 \cdot 0.41H_2O$. The manganese and oxygen atoms are represented by dark grey and red spheres respectively; the potassium atoms and the $H_2O$ molecules are represented by light blue spheres. One unit cell is indicated by the dark grey lines.

**Modulation of the layered structure**

Precession images generated from the raw XRD data show a number of complex features, some of which are depicted in Fig. S9 of the SI. It is apparent that instead of a single spot, there is a cluster of five spots at every integer *hkl* position, which is most obvious in planes perpendicular to the reciprocal *c*-axis such as the *hk0* plane in Fig. S9. These clusters are also visible in a "side-on" perspective, for example in the *h0l* plane. Spots within individual clusters can be indexed assuming a combination of two orthorhombic phases with different lattice parameters and domains formed by small rotations of both phases (see discussion in the SI, section 6). However, although the same clustering of spots was observed for several different crystals from the same batch and for crystals from repeated syntheses, powder XRD shows the existence of only one phase (Fig. S10 and S11 of the SI), hence this multi-domain model can be ruled out. The clustering instead likely arises from a long-period incommensurate modulation as discussed in the SI, section 6.

It was previously proposed using powder XRD that "hybrid structures" can form for certain birnessite compositions, consisting of alternating blocks of orthorhombic and hexagonal structural fragments along the *c* axis that are stacked in an incommensurate manner.[22] In addition, structural modulation might arise due to different occupation of the interlayer space. In general, birnessite compounds can crystallize in different symmetries (triclinic, monoclinic, orthorhombic, or hexagonal) based on the precise stacking of the $MnO_6$ octahedral layers and on the type of cations and the $H_2O$ content in the interlayer space, which can in turn depend on the synthesis conditions.[3,12]

Complex diffraction effects can also arise due to interference involving scattering from different ordered nanodomains. Examples of complicated layered structures with stacking disorder are widely known in clay minerals, giving rise to various 2D diffraction effects. Moreover, various polytypes, different types of defects, and variation in the chemical composition of clay minerals make quantitative analysis of their diffraction patterns challenging.[41] There is often a sequence of ordered/disordered regions referred to as semi-random stacking, where the layers lie randomly but with a limited number of orientations. Therefore, some reflections are not affected by semi-random stacking whereas others produce a two-dimensional diffraction band.[42,43] This can occur for example in turbostratic structures, where some layers are randomly rotated or displaced in the $ab$-plane but the repeating unit is well-defined along the $c$-direction. Therefore, all $hkl$ reflections except for $00l$ form 2D diffraction bands.[42] A similar phenomenon might be an alternative explanation for the spot clustering in K-bir as a result of small layer rotations around the $c$-axis. Although structural disorder in clay minerals has been investigated extensively, there are still many open issues regarding a complete understanding of their structures due to the diversity and variable degree of stacking disorder among individual grains.[44] The structure of K-bir is similarly complex, as revealed by the TEM study discussed below.

## Local structure, stacking faults and charge-ordered nanodomains

TEM images of lamellae cut from a K-bir single crystal demonstrate the layered structure of the compound with well-separated planes of edge-shared $MnO_6$ octahedra. The HAADF-STEM image in Fig. 6a (left part) shows an interface between two different crystal domains, distinguishable by their contrast. The domain in the bottom of the image is perfectly aligned with the zone axis ($a*$) enabling clear visualization of the positions of atomic columns along $a*$. The domain in the upper part of the image is rotated off-zone, resulting in visualization only of the atomic planes ($c$-planes). The corresponding atomic resolution EDS image clearly reveals alternating planes containing manganese and potassium (Fig. 6a, right-hand panel).

DPC-STEM is an emerging technique that has been shown to not only image lighter elements such as oxygen, nitrogen and even hydrogen,[29-31,45] but also the projected electrostatic properties of a thin specimen such as charge density, electrostatic potential and electric fields.[32] In DPC-STEM, a segmented annular detector is used to measure the center-of-mass (COM) of the ronchigram. The position of this COM is a measure of the projected *in-plane* electric field. Using a four-quadrant detector, the COM was approximated by subtracting the signal from the two sets of opposing segments. Thus, the acquired DPC-STEM vector image is a direct representation of the *in-plane* electric field. The integration of this vector image (iDPC) (following the integral form of Gauss' law) represents an electrostatic scalar potential image, and its divergence (dDPC, following the differentiated form of Gauss' law) yields a charge density profile.

Integrated DPC (iDPC) STEM imaging clearly shows adjacent layers of $MnO_6$ octahedra with opposite tilt angles, separated by planes containing K and $H_2O$ columns (Fig. 6b). The lattice parameters ($c$ ~ 13.86 Å, $b$ ~ 5.1 Å) and the local symmetry are consistent with both the average structure determined by SC XRD analysis and the literature data.[10,22,23] It should be noted that the $c$-parameter is slightly smaller than that determined by XRD. The interlayer spacing is sensitive to the $H_2O$ content, and it was previously shown that the loss of lattice water upon sample preparation through FIB and under TEM imaging conditions could lead to a smaller $c$-axis than determined by XRD.[7]

The HAADF-STEM images also reveal the presence of stacking disorder, which might explain the additional reflections in the SC XRD data. In the white rectangle of Fig. 7 (left), the white circles indicate the positions of $MnO_6$ octahedra which are laterally shifted left and right. The solid white lines connect equivalent points in successive layers. In the corresponding fast Fourier transform of this region in Fig. 7 (right-hand panel), a line joining the intense 002 spot with the 022 and $0\bar{2}2$ spots is not perpendicular to the reciprocal c-axis. This implies a monoclinic distortion of the structure with an angle β ≠ 90°. Moreover, there are several streaks of intensity parallel to the reciprocal c-axis at non-integer values of k, which suggests a modulation in the b-direction. Close inspection of the 0kl plane constructed from raw single crystal XRD data reveals distinct spots at slightly different non-integer values of k; since these data were collected at 100 K, the modulation period might be temperature-dependent (Fig. S12 in the SI).

Furthermore, another region with stacking faults is apparent in which there are equal "shifts" of all $MnO_6$ octahedra perpendicular to the layer stacking direction. Three different inter-layer distances can be distinguished in this region, stacked in essentially random fashion (Fig. S13 in the SI).

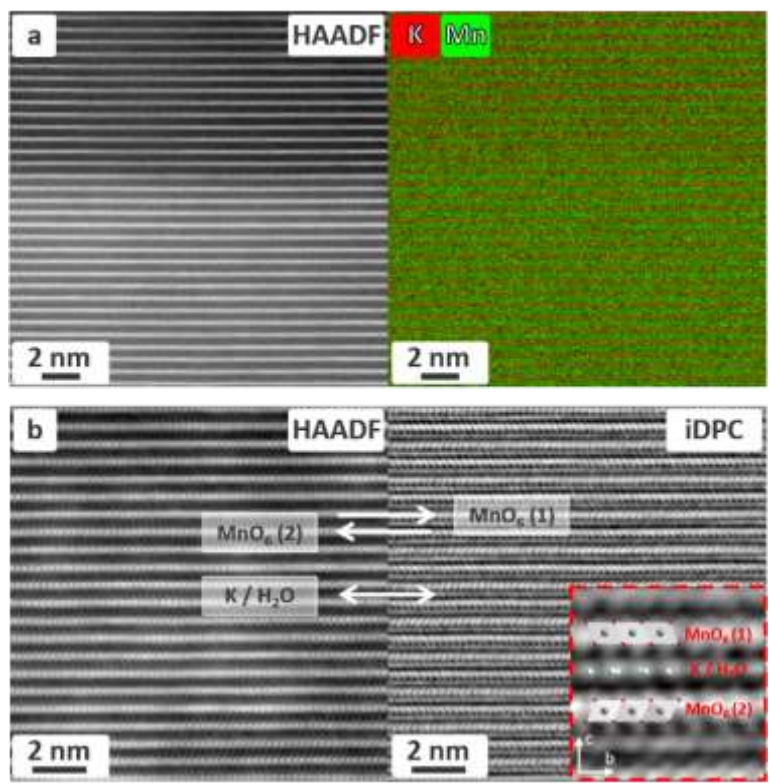

Fig. 6 a) HAADF-STEM image of K-bir and corresponding EDS map showing Mn and K planes perpendicular to *c*-axis; b) HAADF-STEM image of K-bir and corresponding iDPC-STEM imaging revealing the existence of two layers of $MnO_6$ octahedra with opposite tilt angles, separated by layers of K atoms and $H_2O$ molecules.

Moreover, iDPC imaging reveals the presence of some nanoscopic domains (Fig. 8) in which the Mn atomic planes exhibit alternating contrast along the *c*-axis. Such alternating contrast, which is not seen in the HAADF-STEM imaging, originates from periodically alternating electrostatic potentials, and is most likely a result of charge ordering of $Mn^{3+}$ and $Mn^{4+}$ ions along the *c*-axis. These charge-ordered nanodomains (CODs) are potentially important for understanding the origin of the magnetic frustration behavior in these compounds. The minimum size of the CODs can be estimated as 15 x 15 $nm^2$.

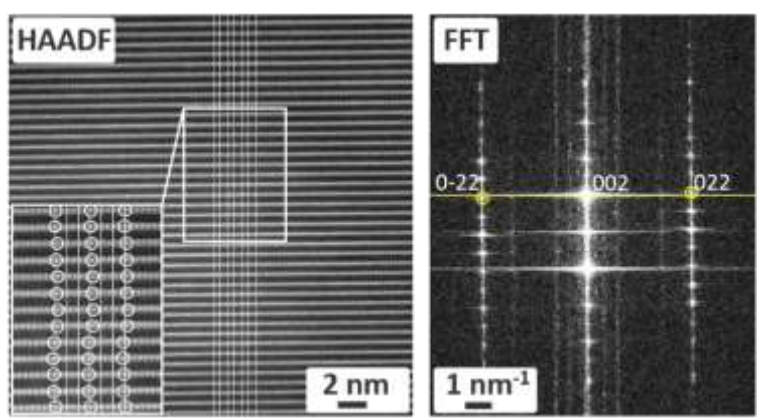

Fig. 7 HAADF-STEM image of K-bir (left) and corresponding FFT of the region outlined by the white rectangle (right).

In addition, it is interesting that the CODs appear to be correlated with regions with a smaller *c*-parameter. The COD in Fig. 8 has a layer spacing of 6.59 Å, and an even smaller spacing of 6.47 Å was determined for another COD with a different contrast pattern (Fig. S14 and S15 in the SI). This compares with a spacing of 6.85 Å for the iDPC image of a representative non-charge-ordered region in Fig. 6b.

We note that the segregation of octahedra containing $Mn^{3+/4+}$ ions was previously demonstrated for K-bir by EXAFS measurements in Ref. 23, but with a different charge ordering model. It was suggested that rows of Jahn-Teller-distorted $Mn^{3+}$ octahedra formed along the *b* axis of the *Cmcm* orthorhombic structure, and were separated by two adjacent rows of $Mn^{4+}$ octahedra leading to a three-fold supercell along the *a* direction.

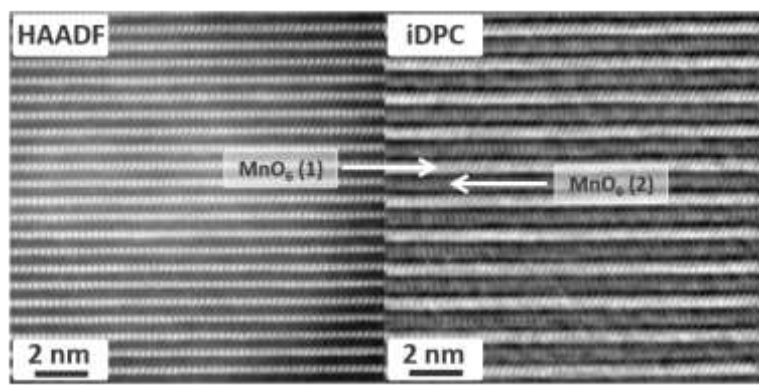

Fig. 8 iDPC-STEM image (right) of a representative nanodomain shows alternating contrast of the $MnO_6$ planes due to charge ordering of $Mn^{3+}/Mn^{4+}$, which is not seen in the HAADF-STEM image (left).

**Influence of structural modulation on magnetic properties**

Understanding the correlation between the structural and magnetic properties is important for the development of possible applications of birnessite compounds. We have recently reported that K/Na-bir, both of which have predominantly antiferromagnetic interactions, exhibit magnetic irreversibility as a function of temperature which originates from the formation of a cluster glass state below the glass freezing temperature. However, we were unable to confirm the origin of these magnetic clusters.[14] Stacking faults involving the arrangement of $MnO_6$ octahedra could certainly enhance the magnetic frustration of the triangular magnetic lattice. Furthermore, we expect that the charge ordered regions of $Mn^{3+}$ and $Mn^{4+}$ ions can have a significant influence on the magnetic properties. First, the Mn-O bond lengths and Mn-O-Mn angles in the $MnO_6$ octahedra are significantly different for $Mn^{3+}$ and $Mn^{4+}$, which will affect the strength and possibly even the sign of magnetic exchange interactions in different regions. Second, $MnO_6$ octahedra containing $Mn^{3+}$ ions, which we observe as planes in the charge-ordered regions, are Jahn-Teller distorted and $Mn^{3+}$ is expected to adopt the high-spin state with antiferromagnetic superexchange in such layers, as in the case of α-$Na_{0.9}MnO_2$ which contains mainly $Mn^{3+}$.[46] However, the presence of $Mn^{4+}$ can suppress the Jahn-Teller distortion and lead to a low-spin $Mn^{3+}$ state in layers that have mixed-valent character.[14] In addition, layers containing only $Mn^{4+}$ can exhibit ferromagnetic $Mn^{4+}$-O-$Mn^{4+}$ superexchange.[47] These different regions with different exchange mechanisms are likely manifested in the magnetic properties as cluster glass behavior.

## Conclusions

Single crystals of K-bir with composition $K_{0.31}MnO_2 \cdot 0.41H_2O$ have been synthesized and investigated. The largest crystals with a size of 250 × 250 × 20 μm³ were grown in a $KNO_3 - B_2O_3$ flux at 700 °C. These crystals exhibit an average orthorhombic symmetry with space group *Cmcm*; the Mn-O bond lengths and Mn-O-Mn bond angles are consistent with mixed-valent $Mn^{3+}/Mn^{4+}$.

A combination of single-crystal XRD together with TEM reveals complex structural features including various types of stacking disorder as well as structural modulation and charge-ordered nanodomains containing alternating planes of $Mn^{3+}$ and $Mn^{4+}$ along the *c*-axis. These ordered $Mn^{3+}/Mn^{4+}$ layers embedded in a mixed-valent matrix can explain the magnetic cluster glass behavior exhibited by K-bir.

Our detailed structural study has revealed interesting features of K-bir that had previously remained hidden. The easily synthesized birnessite family of materials allows tuning of the desired structure by varying the alkali metal and water content, which in turn allows their magnetic properties to be tuned. We envisage a deeper study of their magnetic properties, especially on single crystals, which furthermore should be easily exfoliated to 2D monolayers. The current research contributes to the understanding of these deceptively complex compounds and should stimulate further studies into a potential new class of 2D magnetic materials.

## Author contributions

**Liliia D. Kulish:** Conceptualization, Methodology, Validation, Formal analysis, Investigation, Data Curation, Writing - Original Draft, Visualization, Project administration. **Pavan Nukala:** Validation, Investigation, Data Curation, Writing - Review & Editing. **Rick Scholtens:** Formal analysis, Investigation, Data Curation. **A. G. Mike Uiterwijk:** Formal analysis, Investigation. **Ruben Hamming-Green:** Formal analysis, Investigation, Data Curation. **Graeme R. Blake:** Conceptualization, Methodology, Validation, Investigation, Resources, Data Curation, Writing - Review & Editing, Supervision, Project administration.

## Conflicts of interest

There are no conflicts to declare.

## Acknowledgements

This work was supported by the European Union's Horizon 2020 research and innovation program under a Marie Sklodowska-Curie Individual Fellowship, Grant Agreement Nos. 833550 and 794954. We thank Prof. Ronnie Hoekstra and Dr. Harry Jonkman for valuable discussion and assistance with XPS measurements; Prof. Beatriz Noheda for insightful discussion and resource support; Dr. Václav Ocelik for assistance with the SEM images; Ing. Jacob Baas for technical support; Joshua Levinsky for constructive suggestions during the determination of the crystal structure; Jordi Antoja-Lleonart for valuable recommendations regarding error propagation calculation.